# One-step deposition and in-situ reduction of graphene oxide in glass microcapillaries and application to photonics

*Rodrigo M. Gerosa, Felipe G. Suarez, Sergio H. Domingues, Christiano J. S. de Matos\**

MackGraphe – Graphene and Nanomaterials Research Center, Mackenzie Presbyterian University, 01302-907, São Paulo, Brazil

**ABSTRACT**

Films of graphene oxide (GO) were produced on the inner walls of glass microcapillaries via insertion of a GO water suspension followed by quick drying with a hot finger (no previous surface functionalization required). Individual capillaries from an array could also be selectively GO coated. Raman hyperspectral images revealed the films to be continuous along tens of centimeters. Furthermore, the films could be thermally reduced through an annealing process, which also decreased the concentration of defects. As a proof of principle application, the microcapillaries of photonic crystal fibers were covered with a GO film, leading to the demonstration of fiber polarizers and mode lockers for pulsed fiber lasers. A comparison between GO- and reduced GO-coated fibers revealed that shorter pulses are obtained with the latter.

**INTRODUCTION**

Two-dimensional (2D) materials have attracted great interest since the first experimental isolation of graphene in 2004 [1] with a multitude of applications taking advantage of their unique mechanical, electronic, chemical, thermal and optical properties [2-4]. However, widespread applications will depend upon methods to obtain these materials in large scale [5-7] and, equally important, to transfer them onto the desired substrates. Processes for the latter have been widely studied [8-11] but so far remain a technological bottleneck, especially when curved, microstructured and confined surfaces, such as those of microcapillaries, are to be used as substrates.

Graphene oxide (GO) is possibly the most suitable 2D material for large-scale production, as it is obtained by scalable graphite chemical oxidation methods [12-13]. The several oxygenated functional groups on this material also allow for a good adherence to different substrates, including glass [8] and polyethylene terephthalate (PET)[14]. Additionally, these groups provide reactivity to a number of species, which can be exploited in chemical applications such as catalysis, energy storage (capacitors and batteries) and sensing [15]. Percolation of the GO micro/nanoplatelets allows for a continuous film to be obtained. In addition, GO can be reduced via a number of methods, leading to the conductive material known as reduced graphene oxide (rGO). The disorder that is remnant from the oxidation process, as well as residual hydroxyl groups [16-17], generally make rGO considerably less conductive than graphene. However, the use of microwave pulses has recently enabled conductivities similar to those of chemical vapor deposition graphene to be obtained [18].

The production of GO and rGO films on the inner walls of microcapillaries is of great importance for various applications in two rather diverse fields: microfluidics and photonics. In microfluidics, such a functionalization of the fluidic channels has been exploited for improved gas and liquid chromatography [19-28] and capillary electrophoresis [29-32], but also presents unexplored potential for chemical and bio sensing, as well as catalysis. In photonics, the coating of the inner walls of the capillaries in microstructured optical fibers has been demonstrated [33-35]. The overlap between the guided electromagnetic mode with the nanomaterial has been shown to offer saturable absorption and to produce mode-locking pulses in fiber lasers. Other photonic and optoelectronic applications for such fibers can also be foreseen [36].

In most reported studies, however, a functionalization and/or preparation step was required to allow for the GO/rGO films to attach to the capillaries, thus adding complexity to the process. Also, functionalizing agents can interfere with the light-GO/rGO interaction, affecting its characteristics. In one case, optical forces resulting from a strong electric-field gradient across the capillaries of a microstructured fiber were used to attract and fix the flakes on the surface of the core of the waveguide [36]. However, such a film production method cannot be universally employed for microstructured fibers, as it requires a significant mode overlap with the capillaries, which in turn requires small core diameters. Importantly, the selective coating of only certain capillaries of a microstructured fiber has not yet been reported.

In this paper, we demonstrate a simple, single filling step, process to coat capillaries and microstructured fibers with GO, which is based on the quick evaporation of the liquid host of a GO suspension with a hot finger. This drying method produces a continuous GO film along the whole length of the capillaries, which is firmly attached to the capillary walls. The selective filling of microstructured fibers is also demonstrated. The GO could then be reduced into rGO through thermal treatment under inert atmosphere. As a proof-of-principle application, microstructured fibers coated with GO and rGO are then optically characterized and used as a passive mode locker in fiber laser cavities. It is shown that rGO films yield lower propagation losses and better laser performances.

**FILM PRODUCTION, REDUCTION AND CHARACTERIZATION**

The graphene oxide used in this work was prepared by a modified Hummers method, as described elsewhere [13]. In short, graphite (SP-1 Bay Carbon) was chemically oxidized to produce graphite oxide, a black solid material which was then dispersed in water (1.0 mg/mL) and exfoliated using an ultrasound bath for 5 hours. The bath also ensured a significant fraction of the resulting GO flakes had sub-micron lateral sizes. To prevent clogging of the capillaries, the GO dispersion was filtered (450 nm pore size filter), after which a GO concentration of 75 µg/mL was estimated via ultraviolet absorbance measurements. The resulting suspension was then used to fill the capillaries.

Initial studies were carried out with silica fibers containing one single central capillary, as this geometry makes film inspection and characterization easier. 20-cm long fibers with a nominal 10-µm (125-µm) internal (external) diameter were typically used. For the filling process, the capillary was cleaved and attached to a hypodermic needle using a polymeric UV-curable adhesive (NOA 68T). The needle was connected to a syringe and the suspension GO was pressed into the capillary.

After complete filling, capillary sections were locally heated using a hot finger, with temperatures around 200°C, to induce evaporation of the solvent. The finger was subsequently moved from the extremities towards the center of the fiber, as shown in Figure 1(a). This process was repeated 3 times, and the fiber was then placed in an oven at a temperature of 110° for 5 minutes to remove any remnant liquid residues. This drying method induces minimum evaporation-induced liquid flow that would drag and tend to accumulate the GO flakes at the capillary tips, as in the coffee ring effect [39]. Indeed, drying the capillary directly in the oven did not result in a continuous film along the entire fiber length.

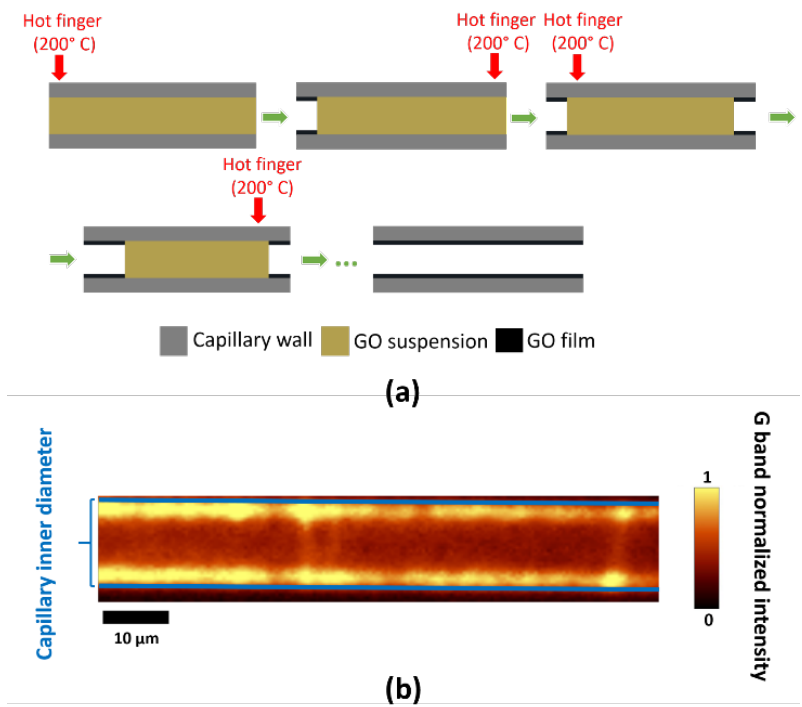

Figure 1. (a) Schematic diagram of the solvent evaporation process inside a microcapillary. (b) Raman map of GO's G band intensity along a longitudinal section of a capillary coated with GO. The blue lines indicate the transverse limits of the capillary.

The characterization of films produced inside the capillary was carried out using a confocal Raman microscope (WITec Alpha 300R; transverse spatial resolution of ~250 nm). The coated fiber rested laterally on the sample stage and the focal plane was adjusted to intersect the capillary's center. Raman spectra were obtained using a 532-nm laser line and Raman maps were obtained by raster scanning the laser across the sample. Figure 1(b) shows the Raman map of a filled capillary in which the intensity of the G band is plotted in color scale. It can be seen that the entire shown section exhibits signal associated with GO, indicating good homogeneity along the capillary axis. The brighter tones close to the walls of the capillary are a consequence of the GO film crossing the focal plane only at these sections (due to the cylindrical geometry). These measurements were repeated along the whole length of the fiber and confirmed a continuous capillary coating. To infer on the adhesion of the GO film on the capillary, a NaOH aqueous solution (0.5 mol/L) was repeatedly flowed through the capillary with no changes detected in the obtained Raman spectra. The good adhesion of GO film to the glass substrate is tentatively explained via the strong interaction (hydrogen bonding) between the hydroxyl and silanol groups present in both materials [37,38].

To obtain rGO, the capillary could then be heated again with the hot finger, while a small flow of nitrogen gas passed through its inside. To establish this flow, the capillary was again fixed to a hypodermic needle, which was connected to a nitrogen pressure system. The Raman spectra of GO in the capillary before and after reduction are shown in Figure 2. The red curve is a Raman spectrum of GO before reduction. It can be seen that the ratio between the D and G bands, $I_D/I_G$, at 1352 and 1610 cm$^{-1}$, respectively, is ~1, which is a characteristic feature of GO. The blue spectrum corresponds to GO after reduction with, as expected, the D and G bands becoming more well defined. However, $I_D/I_G$ increases to 1.30, indicating an increase in disorder in the material lattice caused by the thermal reduction process [40]. To obtain rGO samples with less defects, annealing can be carried out after the reduction. In this process, the capillary fiber was again subjected to an internal nitrogen flow while placed in an oven with the temperature set to 300°C for 2 hours. The green curve in Figure 2 shows the corresponding Raman spectrum, in which $I_D/I_G$ decreased to 0.77.

The demonstrated coating method can be employed to produce functionalized capillaries for the various mentioned applications in microfluidics and photonics. In the latter area, microstructured optical fibers are a particularly interesting and challenging case to study, due to the variety of capillary shapes and sizes, as well as the presence of microcapillary arrays. Here, as a proof of principle experiment, we coat and characterize a range of microstructured fibers. In addition, we show that, by employing previously demonstrated methods [41-42], simultaneous coating of all capillaries or selective coating of individual capillaries can be obtained.

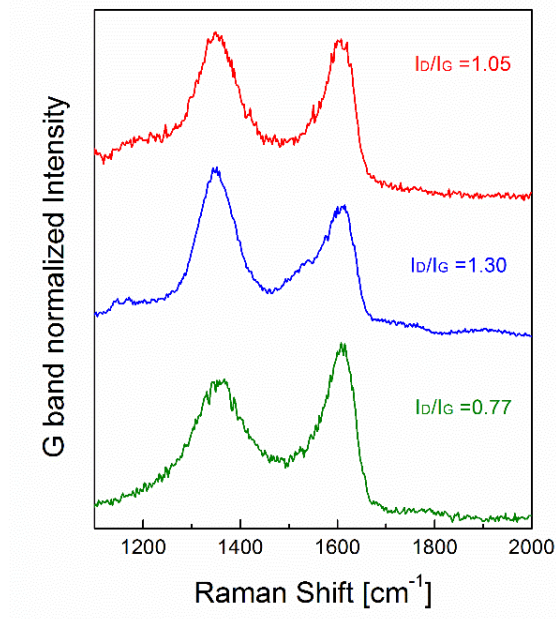

Figure 2. Raman spectra of a GO film inside a capillary before (red) and after (blue) reduction, and after reduction and annealing (green).

**SELECTIVE GO COATING IN CAPILLARY ARRAYS: MICROSTRUCTURED OPTICAL FIBERS**

To demonstrate the simultaneous coating of all capillaries, the photonic crystal fiber (PCF) shown in Figure 3(a) was used. It consists of a ~2.5-μm-diameter solid core surrounded by an array of 2.2-μm-diameter capillaries with a 2.5-μm pitch. The GO suspension and the coating method described in the previous section were used. Figure 3(b) shows a Raman map obtained of the fiber's cleaved face (cross section). The red and blue color bars indicate the intensities of GO's G band and of the silica's 440 cm$^{-1}$ band, respectively. It is possible to note GO's presence in the vast majority of the capillaries. Note that, as is also the case in the other fibers presented here, the GO signal seems to be emerging from the thin glass web structure between the capillaries, rather than from the surfaces of the holes. This is a consequence of the Raman excitation laser being coupled to and guided along optical modes of the web structure, in which case it more efficiently interacts with the GO film on the surface. Figures 3(c) and 3(d) respectively show an optical microscope image and the Raman map of the same fiber, along a longitudinal section. It is evident that the GO is present throughout the shown length.

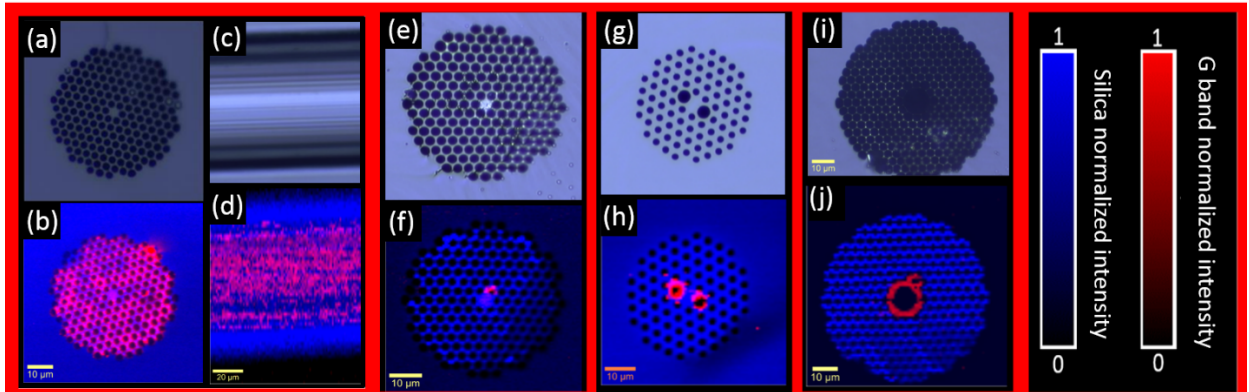

Figure 3. Optical microscopy images (top row) and Raman intensity maps (bottom row – blue: silica's 440 cm$^{-1}$ band; red: GO's G band) of sections of several PCFs. (a),(b) Cross section and (c),(d) longitudinal section of a solid-core PCF. (e),(f) Solid-core PCF with a single coated capillary. (g),(h) Polarization maintaining PCF with the two larger capillaries selectively coated. (i),(j) Hollow-core photonic bandgap fiber with a selectively coated core.

For selectively filling individual capillaries, a technique demonstrated by Gerosa *et al.* [41] was used. It uses a micropipette with an outer tip diameter of ~1 μm fixed to a triaxial micropositioner. With the guidance of a microscope, the pipette is inserted into the desired hole and the liquid is inserted upon pressure application. This setup is best described in the Supplementary Information. As a demonstration of the ultimate flexibility of the method, one single capillary next to the solid core of the previously described PCF was GO coated, as shown in Figures 3(e) and (f). Such a geometry may be exploited for efficient light-nanomaterial interaction [43].

Alternatively, the larger capillaries of microstructured optical fibers presenting a nonuniform microstructure can be selectively coated. For this purpose, the glass near both tips of the fiber is heated and softened so that the smaller capillaries collapse [42] prior to the filling with the GO suspension. Here, this method was employed to coat both a polarization maintaining PCF and a hollow-core PCF. The polarization maintaining PCF cross section is shown in Figure 3(g). It consists of two 4-µm diameter capillaries in either side of a solid core, which are surrounded by an array of 2-µm diameter capillaries. The resulting Raman intensity map is shown in Figure 3(h), from which it can be seen that only the two larger capillaries were coated, resulting, as will be shown in the next section, in a polarization dependent loss. The cross section of the hollow-core PCF is shown in Figure 3(i) and consists of a 10-µm diameter core surrounded by a 3.5-µm diameter capillary array with a 3.8-µm pitch. The Raman map of the coated fiber is depicted in Figure 3(j) and shows selective filling of the core and one adjacent capillary.

## APPLICATION OF A COATED MICROSTRUCTURED FIBER: INTEGRATED POLARIZER AND LASER MODE LOCKER

Obtaining continuous GO films on the inner walls of microstructured fibers adds functionality to these waveguides, allowing them to be used as, e.g., integrated polarizers or passive intracavity pulse shapers for mode-locked lasers [33-35]. Here, we infer the polarizing power and the mode locking performance of one of the coated PCFs. The polarization maintaining PCF is chosen for this purpose, for being designed to be used at the 1550-nm wavelength (in the telecommunications' C band) and for allowing for a low splicing loss with a standard single-mode telecommunications fiber (SMF-28). This fiber also has the advantage of presenting a high birefringence and, thus, nondegenerate polarization eigenmodes with negligible cross coupling. This last feature means that polarizing fibers can be obtained by GO coating only the two larger holes (Figure 3(h)), which induces different losses on the two eigenmodes.

Two coated PCF samples were prepared following the methods described in the previous sections, one of which also underwent the reduction and annealing steps. Raman spectra and Raman intensity maps of the samples are shown in the Supplementary Information. The samples had an initial length of 10 cm and were spliced to standard telecommunications patchcords with FC/APC connectors to facilitate integration with other fiber components and incorporation into an Erbium-doped fiber laser cavity. Each splice had a typical loss of ~1.1 dB.

The samples were optically characterized for linear loss and polarization extinction ratio using an experimental setup (shown in the Supplementary Information) consisting of a tunable laser (1510 to 1640 nm), a Glan-Thompson cube polarizer and a halfwave plate. For loss measurements, the fibers were cutback and their transmission loss (at the polarization with the highest loss) evaluated as a function of length, as shown in Figure 4. In the figure, dots and lines are experimental data and linear fits, respectively. The red and green data are for the GO sample and the rGO sample, respectively. The good fits indicate that the loss is homogeneously distributed along the fibers, which again confirms the continuity of the formed films. It is evident that the rGO sample presents a lower propagation loss, determined to be 3.5 dB/cm, while the GO sample exhibited a propagation loss of 4.1 dB/cm. Note that the green triangle in Figure 4 represents the loss obtained with the rGO sample before the reduction and indicates that, under these circumstances, both samples present very similar losses (26.9 and 26.1 dB). This result points to the good reproducibility obtained with the demonstrated coating method. The loss decrease with the reduction of GO may be accounted for by the fact that the reduction decreases the distance between the exfoliated platelets, making the rGO film smoother and, therefore, less scattering [44-46].

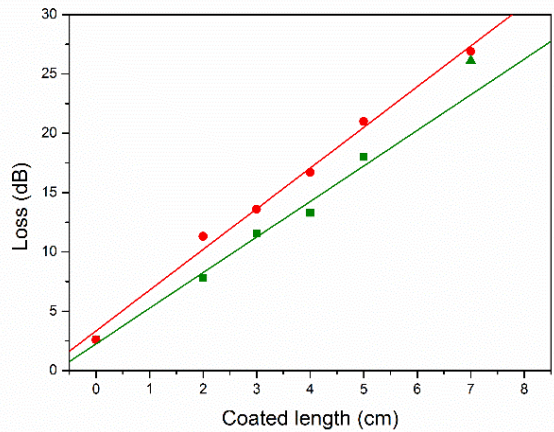

Figure 4. Transmission loss as a function of PCF length for the GO (red) and rGO (green) samples. Dots are experimental data and lines are linear fits. The green triangle was measured in the rGO sample prior to GO reduction.

Figure 5 shows the polarization dependence of the power transmitted through both samples (for a 5-cm length), which was measured by rotating the input linear polarization in 20º steps. It is possible to notice an extinction ratio, between the maximum and minimum transmission, of 7.3 dB for the rGO sample and of 5.9 dB for the GO sample. Note that for the lowest transmission axis the loss value for the two fibers is very similar; the largest difference is found for the axis of maximum transmission, which also points to the decrease in additional (scattering) losses by the GO reduction process.

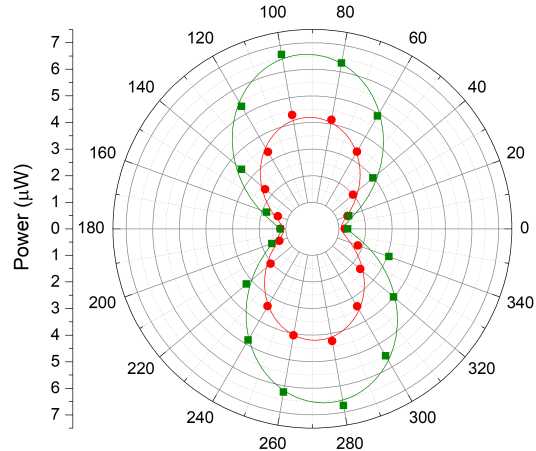

Figure 5. Transmitted power as a function of the incident polarization angle for the GO (red) and rGO (green) samples (5-cm fiber length).

The coated fiber samples were then incorporated in a fiber ring laser cavity, to be tested as a mode locker. The experimental setup is detailed in the Supplementary Information and consists of a 15-cm-long erbium doped fiber, a semiconductor pump laser at 980 nm, a 980/1550 nm WDM coupler with a built-in isolator for 1550 nm, an in-line fiber polarization controller, and a 30% output coupler.

The laser cavity was tested with the GO and rGO samples with lengths of 2, 3, 4 and 5 cm, the insertion losses of which are shown in Figure 4. For a PCF length of 5 cm, lasing was not achieved due to the high losses (21 and 18 dB for the GO and rGO samples, respectively, in Fig. 4). For a 4-cm length (losses of 16.7 and 13.3 dB), mode locking was not obtained, but the laser operated in the continuous-wave (CW) regime, centered at 1530 nm. For a length of 3 cm (losses of 13.6 and 11.6 dB) CW-only operation was obtained with GO (at 1530 nm), but mode-locking operation with the output centered at 1550 nm was attained with the rGO sample. However, the spectral bandwidth of the pulses was still low (0.1 nm FWHM), indicating relatively long pulses.

Finally, for a length of 2 cm (losses of 11.3 dB and 7.8 dB), mode-locking operation was obtained with the GO sample, with a central wavelength of 1550 nm and a bandwidth (FWHM) of ~0.1 nm (Figure 6(a)), as well as with the rGO sample, with a central wavelength of 1561 nm and a bandwidth (FWHM) of 4 nm (Figure 6(c)). The pulse trains obtained with the GO and rGO samples can be seen in Figures 6(b) and 6(d), respectively. They correspond to 64 and 48.2 MHz repetition rates, respectively, which match the cavity round trip frequencies for each cavity.

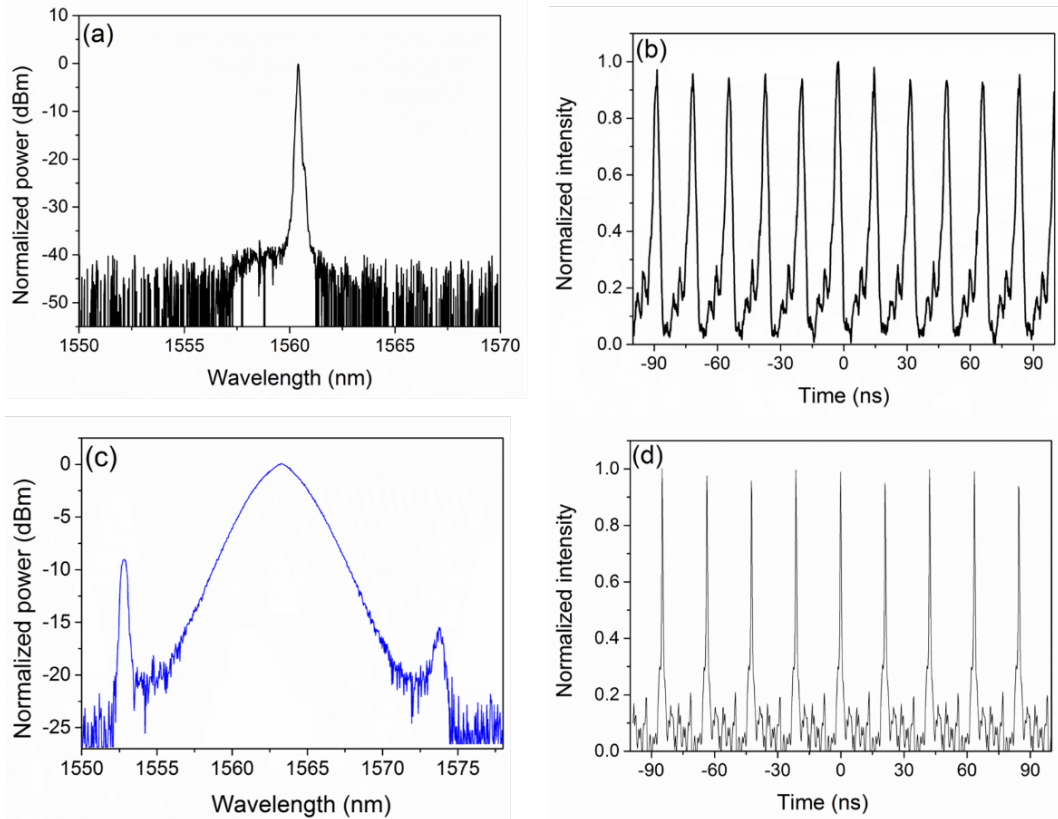

Figure 6. (a),(c) Optical spectra and (b),(d) pulse train temporal profiles of the mode-locked lasers containing (a),(b) the GO- and (c),(d) the rGO-coated fibers.

As can be seen, the rGO fiber sample with a 2-cm length yielded the best mode locking performance, leading to the pulses with the broadest bandwidth. This result apparently contradicts that of Sobon et al. [47], who have shown GO and rGO mode lockers to present similar performance. However, in that case, planar GO and rGO films on a silica substrate were used, and were traversed by the intracavity laser beam. The better performance presented here for the shortest rGO sample is likely to be a consequence of the lower obtained losses. Also, the mode-locking mechanism is likely to be associated not only with the 2D materials' saturable absorption, but also with nonlinear polarization rotation, which benefits from the sample's polarization dependent loss. The ultrafast response of this mechanism may help promoting the larger observed bandwidth.

## CONCLUSIONS

A simple, single step, and efficient method to coat the inner walls of microcapillaries with graphene oxide and reduced graphene oxide was demonstrated. It is based on the insertion of a graphene oxide water dispersion into the capillaries with a subsequent quick drying of the water, which is shown to produce an oxide film that is continuous and presents reproducible characteristics. Reduced graphene oxide coating is obtained via heat treatment of the oxide film under nitrogen flow. As a proof of principle application, the microcapillaries of several photonic crystal fibers were coated, with selective coating of specific capillaries in an array also demonstrated. The optical characterization of a graphene oxide and reduced graphene oxide coated polarization maintaining photonic crystal fiber was then presented, with the reduced oxide film leading to lower propagation losses. A 2-cm-long fiber with this film was then used to produce 4-nm bandwidth mode-locked pulses at the fundamental repetition rate of a ring fiber laser. The coating method may also be used to produce samples for other applications, such as chromatography or photocatalysis, as well as for the coating of capillaries with other chemically exfoliated 2D materials.

## AUTHOR INFORMATION

**Corresponding Author**
* cjsdematos@mackenzie.br

## ACKNOWLEDGMENTS

This work was funded by FAPESP (SPEC Project 2012/50259-8 and Thematic Project 2015/11779-4), Conselho Nacional de Desenvolvimento Científico e Tecnológico (CNPq), and Fundo Mackenzie de Pesquisa.